# Dark Matter

Douglas M. Snyder
Los Angeles, California[1]

Dark matter has been an issue of concern for some time. The evidence for dark matter has over the years become increasingly strong and the problem appears in a variety of contexts (e.g., on the level of a single galaxy, a large cluster, the universe).[1,2,3,4] For example, Carr wrote:

> The evidence for dark matter, on all scales from star clusters ($10^6$ solar masses) to the Universe itself ($10^{22}$ solar masses), has built up steadily over the past 50 years...Although the strength of the evidence on different scales varies considerably, thre is now little doubt that only a small fraction of the mass of the Universe is in visible form. (Carr, 1994, 531)[2]

There is a particular approach to the problem of dark matter that has not been explored, one which from the point of view of general relativity is logically valid.

## The Dark Halo

In this paper, this alternative approach to the problem of dark matter will be discussed in terms of the dark halo for a single galaxy. Yet the general argument to be made may be applicable to other contexts where significant evidence exists for dark matter. It is known that a galaxy is often associated with a large rotating body of cold hydrogen gas that often reaches beyond the galaxy's visible limit. The rotational velocity of this gas circling the core of the galaxy can be used to derive the the galaxy's attractive gravitational field.[3] From the derived gravitational field, an estimate of the mass associated with this field can be found easily and the gravitational masss is reasonably considered to be the gravitational mass of the galaxy. The dark halo of a single galaxy can be determined by using the following relation that is derived in Newtonian mechanics:

$$GM/R \propto v^2 \:. \tag{1}$$

where $G$ is the gravitational constant, $M$ is the mass of the attracting object, $R$ is the radial distance from the center of mass of the attracting object, and $v$ is the magnitude of the tangential velocity of the attracted rotating hydrogen gas.[1] The proportionality holds because $GMm/R^2$ is the attractive gravitational force

# Dark Matter

on the hydrogen gas and $v^2/R$ represents the centripetal acceleration of the gas due to this gravitational force, where $m$ is the gravitational mass of the attracted object.

A dark halo is indicated because the gravitational field derived in this manner indicates the presence of much more mass than is visible. In the case of single galaxies, in particular spiral galaxies, as $R$ increases it has been found that the tangential velocity $v$ of the rotating hydrogen gas remains the same.[1,3] Thus $M$ must increase as well according to the proportionality noted above. The visible mass of a galaxy does not match the amount of mass determined with this proportionality. Indeed, it is many times less, generally considered to be about 10% of the mass determined using the proportionality noted above.[1] Wyse and Gilmore wrote:

> The mass of a galaxy appears to be distributed very differently to the starlight, and to be located in a more spatially extended "dark halo". The total dark halo mass is perhaps 10 times larger than that associated with all the stars in the galaxy, together with the identified gas. (p. 40)

Various theories have been put forward to account for this missing mass. Generally, these theories either propose that the dark matter is baryonic in nature or is a type of exotic elementary particle. The baryonic alternatives for the dark halo of a galaxy include smaller low-intensity galaxies or black holes.

## An Earlier Situation

A situation confronting physicists from an earlier time might be helpful in dealing with dark matter. The resolution of the earlier problem involved accepting the empircal results as they were and adopting a straightforward theoretical model that was consistent with them, instead of holding to a model to which significant alterations had to be made to predict the empirical results.

In the late 1800s and early 1900s, physicists looked for the medium through which light traveled. They knew that light was electromagnetic raidation and thus wave-like in nature. But physicists could find no evidence for a medium, the ether, through which the light traveled. Chief among this evidence was that the velocity of light was independent of the velocity of the body emitting the light. Physicists, though, interpreted their empirical results so that they could continue to maintain that the ether existed.[5] These interpretations essentially involved considering the velocity of light to be dependent on the velocity of the emitting body and changing other parameters to

# Dark Matter

account for the empirical result that the velocity of light is invariant.

In 1905, Einstein proposed that the invariant velocity of light in inertial reference frames traveling at uniform translational velocities relative to one another be accepted as such and, further, that the ether did not exist. On these premises and the additional one that the laws of physics are the same in inertial reference frames in uniform translational motion relative to one another, he built the special theory of relativity.

Is it possible that physicists face a similar situation in that they are looking for dark matter and it does not exist? It is specifically gravitation that physicists believe indicates the presence of dark matter that they need to account for. Conceptually there is another way to account for the "extra" gravity without introducing mass to account for it.

## Reference Frames and Gravitation

In 1916, when Einstein proposed the general theory of relativity, he did so precisely because he could not find evidence for the privileged role granted to inertial reference frames in the special theory. Einstein also came to understand that an inertial reference frame experiencing a uniform gravitational field is equivalent to a uniformly accelerating reference frame. Einstein was able to develop a metric for spacetime for an inertial reference frame experiencing a uniform gravitational field by considering the uniformly accelerating reference frame as a series of local Lorentz reference frames with adjacent frames moving at slightly different uniform translational velocities relative to one another. For these Lorentz reference frames, the spatiotemporal relationships developed for inertial reference frames in uniform translational motion relative to one another developed in special relativity held. Generally, the gravitational field associated with an inertial reference frame is associated with mass, and Einstein thus arrived at a new understanding of gravitation.

A second way to account for the gravity indicating the presence of "dark matter" is in terms of the pattern of reference frames alone that are used to develop the spacetime structure that in the general theory constitutes the gravitational field. The gravitational mass that in Newtonian mechanics is unavoidably associated with gravitation is in the general theory not central to the development of spacetime curvature.

Though an unusual suggestion, the thesis that the gravitational field generally associated with dark matter is actually due to the reference frames themselves used to develop spacetime curvature is just an extension of

# Dark Matter

Einstein's own principle of equivalence concerning uniformly accelerating reference frames and inertial reference frames experiencing a uniform gravitational field. Einstein wrote of situations where an observer would maintain that he was in an inertial reference frame experiencing a gravitational field whereas to an observer "outside" the former observer's reference frame, the former observer would be seen to be in an accelerating reference frame.[4,5] For example, he discussed a man in a windowless room in deep space. The room is towed by a rope attached to one end so that the room accelerates at a uniform rate. What does the man inside the room experience? He feels as if he is in a gravitational field. For example, if the man in the room hangs a rope from the ceiling and attaches an object to the bottom, the rope hangs toward the floor. What does the man in the room think? He thinks, "The suspended body experiences a downward force in the gravitational field, and this is neutralised by the tension of the rope" (Einstein, 1917/1961, p. 68). What does an observer outside the room deep in space think? He thinks, "The rope must perforce take part in the accelerated motion of the chest [the room], and it transmits this motion to the body" (Einstein, 1917/1961, p. 68).[4]

A passage from Misner, Thorne, & Wheeler, 1973 can help make the present suggestion more explicit.

> A tourist in a powered interplanetary rocket feels "gravity." Can a physicist by local effects convince him that this "gravity" is bogus? Never, says Einstein's principle of the local [over a small area] equivalence of gravity and accelerations. But then the physicist will make no errors if he *deludes* [italics added] himself into treating true gravity as a local illusion caused by acceleration. Under this delusion, he barges ahead and solves gravitational problems by using special relativity: if he is clever enough to divide every problem into a network of local questions, each solvable under such a delusion, then he can work out all influences of any gravitational field. (p. 164)

In the suggested theory, the physicist is not deluding himself by treating "true" gravity in terms of accelerating reference frames. This is the meaning of "true" gravity.

## Conclusion

Gravitational force manifested in its affect on rotational velocity is what indicates the presence of dark matter in individual galaxies. Newtonian

# Dark Matter

mechanics is generally used to derive the relationship between rotational velocity and gravitational force. In Newtonian mechanics, mass is associated with gravitation. At the same time, in such galaxies there appears to be insufficient gravitational mass to account for the size of the gravitational force deduced from rotational velocity. The general theory of relativity encompasses Newtonian mechanics and also indicates that reference frames provide the means to develop spacetime curvature that expresses the gravitational field. It is reasonable to conclude that because of the more fundamental character of the general theory of relativity, the reference frames that are a key element of the general theory form the basis for understanding gravitation, and not mass.

## Endnotes

[1] Email: dsnyder@earthlink.net

[2] The term "solar masses" is used in the quote whereas in the article by Carr the graphic symbol for this term was used.

## References

1. V. Trimble, Existence and Nature of Dark Matter in the Universe, *Annu. Rev. Astron. Astrophys.*, *25*, 425 (1987).
2. B. Carr, Baryonic Dark Matter, *Annu. Rev. Astron. Astrophys.*, *32*, 531 (1994).
3. R. Wyse & G. Gilmore, What's the Matter?, *Physics World*, *8*(6), June, 1995, p. 39.
4. Bothun, G. D., The Ghostliest Galaxies, *Scientific American*, *276*(2), February (1997), p.56.
5. R. Resnick, *Introduction to Special Relativity*. (John Wiley & Sons, 1961) p.??????.
6. C. W. Misner, K. S. Thorne, and J. A. Wheeler, *Gravitation* (W. H. Freeman, 1973), p. 164.
7. A. Einstein, "On the electrodynamics of moving bodies," in H. Lorentz, A. Einstein, H. Minkowski, and H. Weyl (editors), *The Principle of Relativity, a Collection of Original Memoirs on the Special and General Theories of Relativity* (Dover, 1952, originally 1905), p. 53.
8. A. Einstein, *Relativity, the Special and the General Theory* (Bonanza, 1961, originally 1917), p. 44.
9. A. Einstein, "The foundation of the general theory of relativity," in H. Lorentz, A. Einstein, H. Minkowski, and H. Weyl (editors), *The Principle of Relativity, a Collection of Original Memoirs on the Special and General Theories of Relativity* (Dover, 1952, originally 1916), p. 113.
10. A. Einstein, *The Meaning of Relativity* (Princeton University Press, 1956, originally 1922), p. 57.